\documentclass[twocolumn]{aastex62}

\usepackage{amsmath}	
\usepackage{amssymb}	

\newcommand\aastex{AAS\TeX}
\newcommand\latex{La\TeX}

\usepackage[normalem]{ulem}

\newcommand{\pavg}{\langle p \rangle}

\newcommand{\beq}{\begin{equation}}
\newcommand{\seq}{\end{equation}}

\newcommand{\gv}[1]{\ensuremath{\mbox{\boldmath$ #1 $}}} 

\graphicspath{{./}{figures/}}

\received{February 27, 2019}
\revised{March 18, 2019}
\accepted{March 22, 2019}
\submitjournal{ApJL}

\shorttitle{cloud coalescence} 
\shortauthors{Waters \& Proga}

\begin{document}

\title{Cloud coalescence: a dynamical instability affecting multiphase environments}

\correspondingauthor{Tim Waters}
\email{waters@lanl.gov}

\author{Tim Waters}
\affil{Theoretical Division, Los Alamos National Laboratory}

\author{Daniel Proga}
\affiliation{Department of Physics \& Astronomy, University of Nevada, Las Vegas} 





\begin{abstract}
The mass and size distributions are the key characteristics of any astrophysical objects, including the densest clumps comprising the cold phase of multiphase environments.
In our recent papers, we showed how individual clouds of various sizes form and evolve in AGN.
In particular, we showed that large clouds undergo damped oscillations as a response to their formation process.  
Here we followup this investigation, addressing how different size clouds interact.  
We find that smaller clouds become trapped in the advective flows generated by larger clouds.
The explanation for this behavior leads to a rather remarkable conclusion: 
even in the absence of gravity, complexes of clouds are dynamically unstable. 
In an idealized environment (e.g., one free of turbulence and magnetic fields) a perfectly symmetric arrangement of static clouds will remain static,
but any small spatial perturbation will lead to all clouds coalescing into a single, large cloud, given enough time.  
Using numerical simulations, we investigate the main factors that determine the rate of coalescence.  
Besides the cloud separation distance, we find that the transient response of clouds to a disturbance is the primary factor.  
Turbulent motions in the flow can easily suppress this tendency for spatially well-separated clouds to coalesce, so it is as yet unclear if this phenomenon can occur in nature.  Nevertheless, this work casts strong doubts on a recent hypothesis that large clouds are prone to fragmentation.

\end{abstract}

\keywords{hydrodynamics --- instabilities --- radiation: dynamics --- (galaxies:) quasars: general}


\section{Introduction} \label{sec:intro}
The parsec and subparsec-scale environments of active galactic nuclei are inferred to host multiphase structures, namely the obscuring `dusty torus' and the broad and narrow line emission regions (for recent reviews, see e.g., Netzer 2015; Padovani et al. 2017; Almeida \& Ricci 2017; Hickox \& Alexander 2018).  
Because the cooling times of the cold phase gas in these dense environments can be very short (on the order of hours to days in the broad line region), the individual clouds may be subject to the dynamical instability identified herein on timescales that can be directly observed.  

In this letter we show that (i) multiple clouds interacting in the nonlinear regime of thermal instability (TI; Field 1965) tend to coalesce and (ii) that this occurence is not unique to clouds formed via TI but rather is a generic property of multiphase gas dynamics.   

We refer to cloud coalescence as a dynamical instability because we find that the only way for multiple interacting clouds to reach a steady state is if the cloud spacings are perfectly symmetric, and even then, a small displacement from this state will cause the clouds to merge.   
This process, if left unchecked in an idealized (e.g., non-turbulent and unmagnetized) cloud complex, would inevitably lead to large clouds, i.e. ones with characteristic cloud sizes $d_c$ significantly exceeding the local `acoustic length' of the gas,
\beq \lambda_c = (c_s t_{cool})_c,\seq 
where $c_s$ is the adiabatic speed of sound and $t_{cool}$ is the cooling time (defined as the ratio of the gas internal energy, $\mathcal{E} = c_{\rm{v}} T$, and the cooling rate $\Lambda$ in units of $\rm{erg\,g^{-1}\,s^{-1}}$).  Here, the subscript notation denotes the evaluation of quantities at the stable cold phase, which has mass density $\rho_c$ and temperature $T_c$.  In other words, cloud coalescence naturally leads to the non-isobaric regime of gas dynamics, in which  
$d_c/c_{s,c} = t_{dyn}>> t_{cool}$, 
implying significant deviations from pressure equilibrium within such multiphase systems.  

A recent hypothesis that has gained a lot of attention is the notion that clouds in this regime
are prone to fragmentation --- the opposite of coalescence.  Namely, McCourt et al. (2018; hereafter M+18) speculated that a large cloud may restore pressure equilibrium on short (dynamical) timescales by `shattering' into many tiny `cloudlets', each with a characteristic size $\lambda_c$.  

We recently uncovered the dynamics of newly formed non-isobaric clouds in 1D (Waters \& Proga 2018; hereafter `Paper~1').  In the context of this study, we highlight two findings from that work: 
(i) non-isobaric clouds are `content' to remain large.  Rather than through some rapid fragmentation process, large clouds formed from TI regain pressure equilibrium on timescales long compared to the dynamical time by undergoing damped oscillations (in size, density, and temperature).
(ii) Larger clouds require larger velocity fields to maintain their structure \emph{when they are undergoing oscillations}.  Here we build upon this study to understand non-isobaric behavior in both 1D and 2D when multiple clouds interact.
Our results are presented in \S{2} and \S{3}, followed by a discussion and our conclusions in \S{4}. 
 
\begin{figure}
\includegraphics[width=0.45\textwidth]{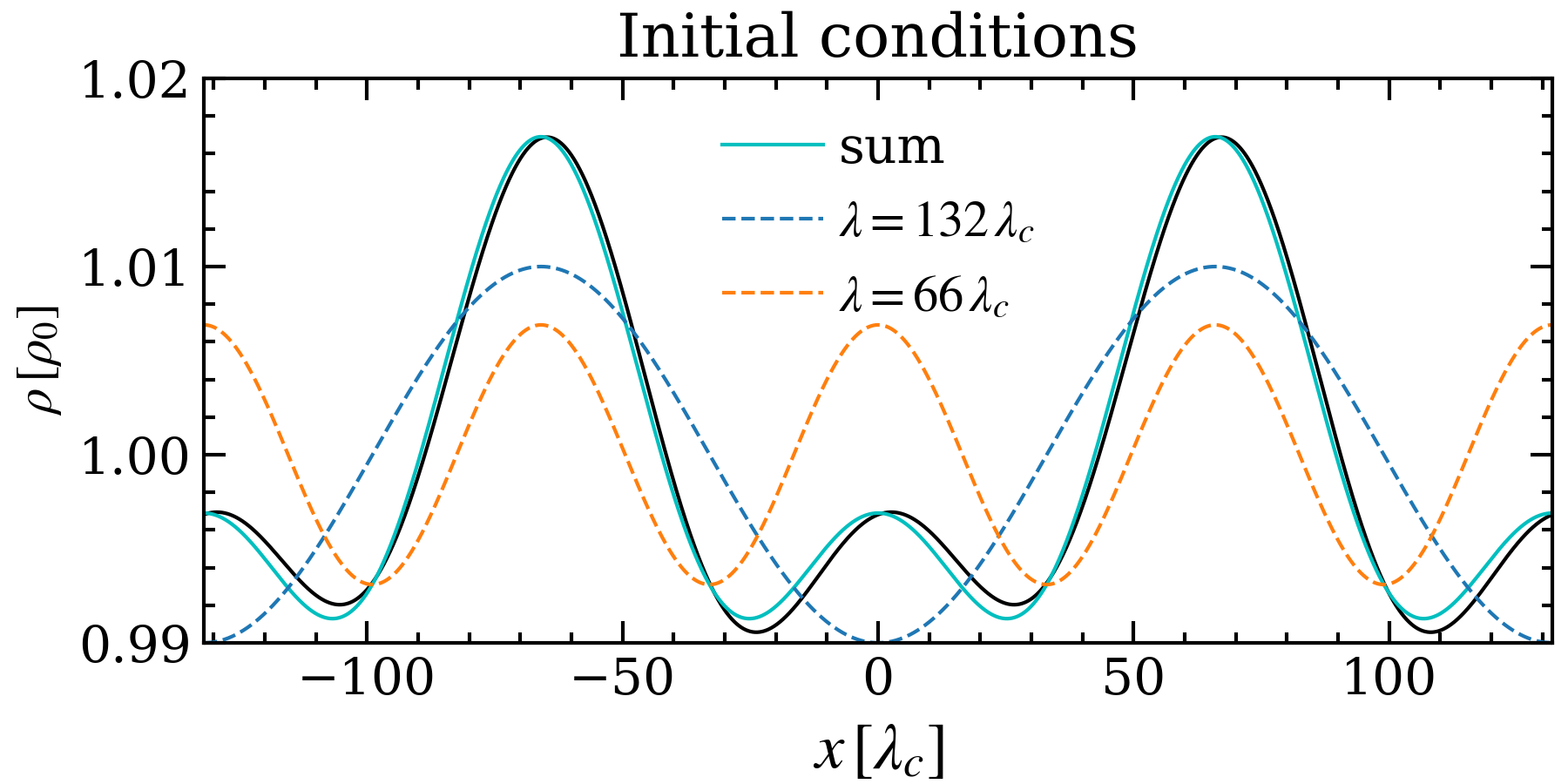}
\includegraphics[width=0.45\textwidth]{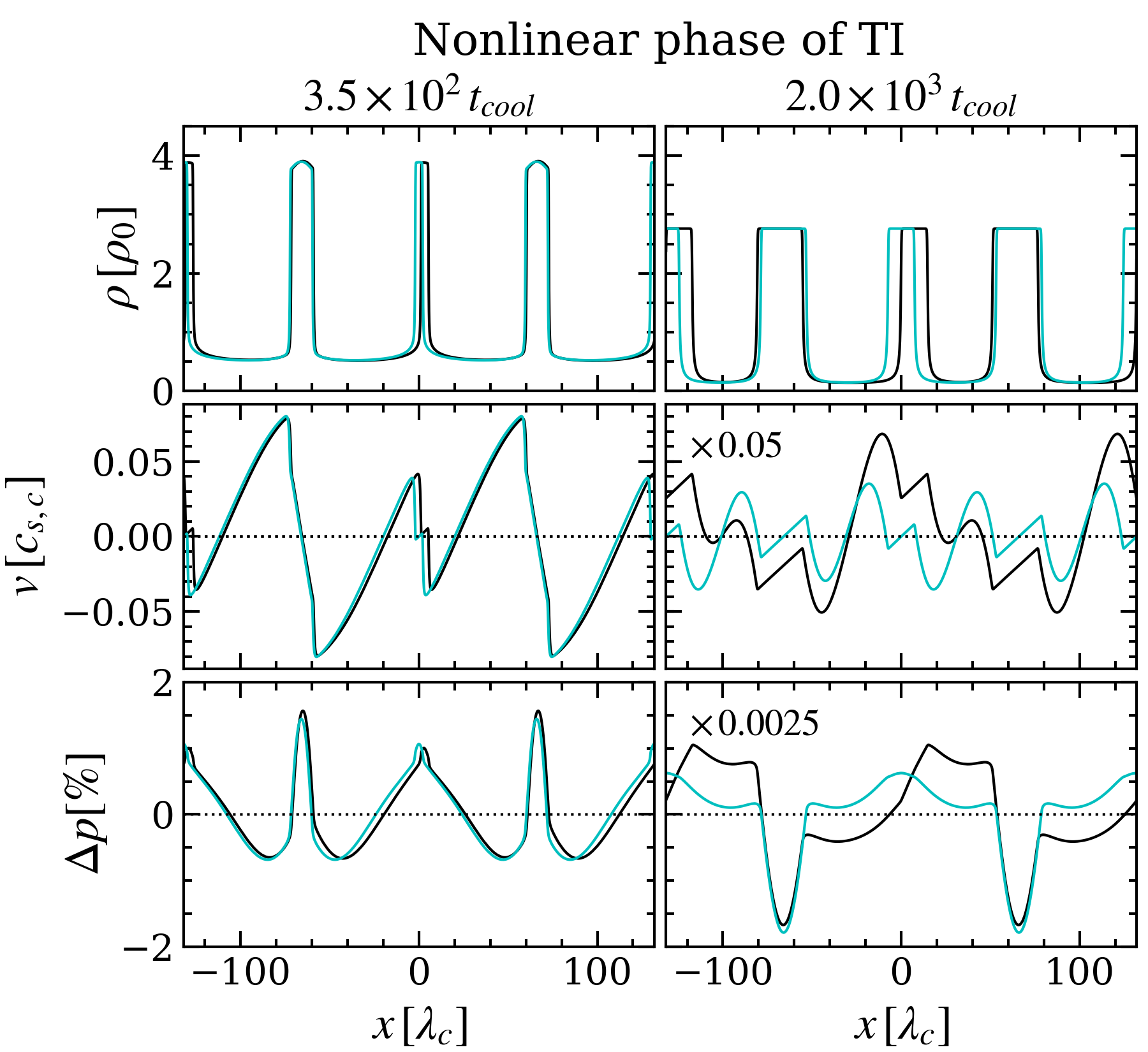}
\caption{Initial conditions and early nonlinear evolution of our two 1D TI runs, Run A (black curves) and Run B (cyan curves).  Top panel: our initial density profile consists of a superposition of two modes with wavelengths $\lambda = 66\,\lambda_c$ and $\lambda = 132\,\lambda_c$, and with mutual amplitudes and phase shifts chosen so that 4 clouds will form in a domain with size $x \in [-132,132]$.
The phase shift for Run A is slightly different than $\pi$, and this small asymmetry results in coalescence. 
Bottom panels: 
Profiles of the density and velocity (top and middle panels), as well as the pressure percent differences (bottom panels, where $\Delta p = 100\times(p - \pavg)/\pavg$, with $\pavg$ denoting the domain average of $p$) at the early saturation phase of TI (left column) and much later (right column) when Run B has almost reached a steady state.  Notice the difference in the $\Delta p$ profiles in the bottom right panel; Run A has an asymmetric profile indicative of unbalanced forces.   In this panel and the one above, we zoomed in by the reciprocal of the numbers in the top left corners to make the profile shapes visible. 
} 

\end{figure}

\begin{figure*}
\includegraphics[width=0.99\textwidth]{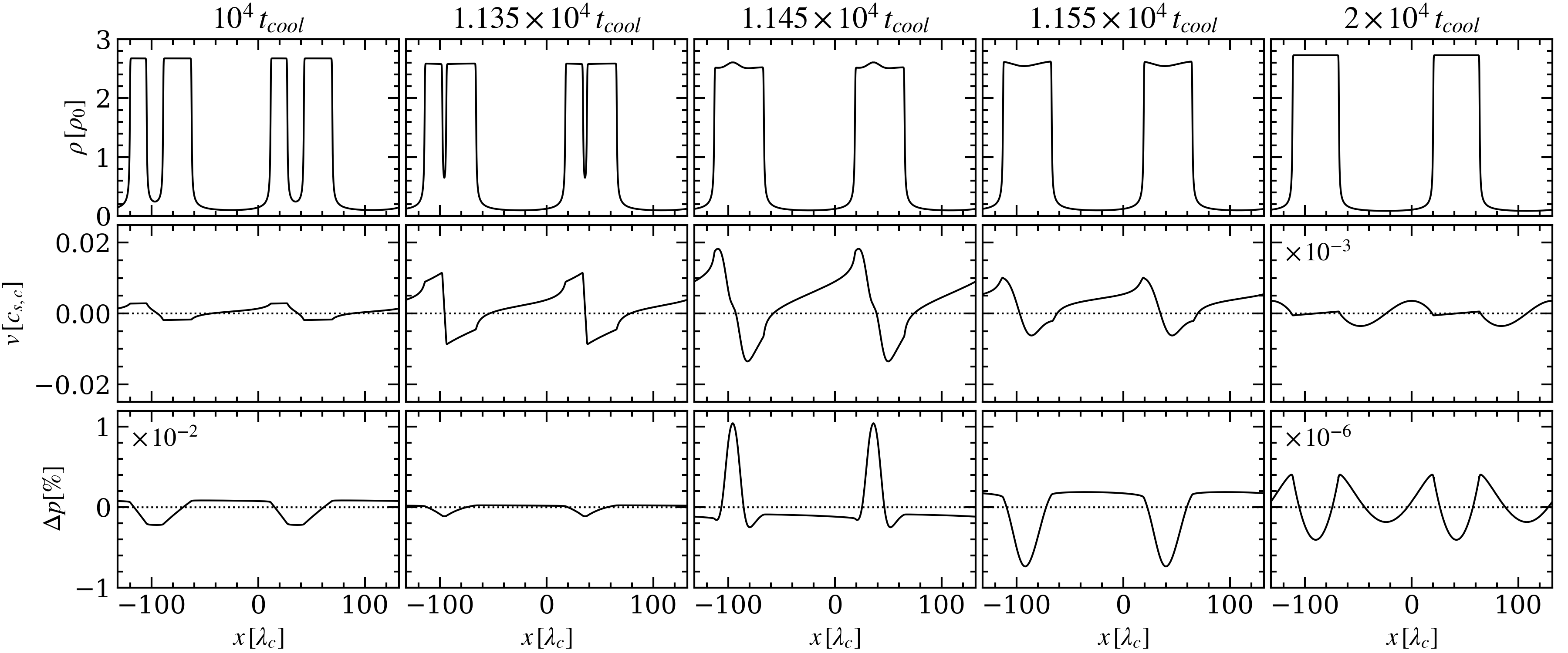}
\caption{
Similar plots as the bottom panels in Fig.~1 but only Run A is shown and for times shortly prior to (left column), during (middle three columns), and long after (right column) the coalescence process.  Notice that deviations from pressure equilibrium are only about $0.001$\% before coalesence and less than $10^{-6}\%$ as a steady state is reached.  
} 

\end{figure*}

\section{TI simulations}
In Paper~1, we studied the dynamics of individual non-isobaric clouds through the nonlinear regime and into a steady state.  Here we present similar 1D simulations using Athena++ (Stone et al. 2008 and in prep.) for the same cooling function (that of Blondin 1996) but with two superimposed perturbations as initial conditions (ICs) instead of a single eigenmode of TI, as shown in the top panel of Fig.~1.  
All of our runs have a fixed resolution of 16 zones/$\lambda_c$, sufficient to fully resolve cloud interfaces (see Proga \& Waters 2015), which have a characteristic width $\lambda_F = 3.07\,\lambda_c$ (as derived from the Spitzer value of the initial equilibrium state; see eqn. 3 in Paper~1), where $\lambda_F$ is the Field length.\footnote{This work uses different units than Paper~1, where it was natural to define quantities with respect to the initial conditions.  The relation is $(\lambda_{th},t_{cool})_{\text{Paper\,1}} = (16.5\,\lambda_c, 9.9\,t_{cool})$.} 
To demonstrate how cloud coalescence occurs while at the same time conveying our viewpoint that this process can be considered a dynamical instability, we compare two runs: 
(i) Run A (asymmetric ICs), in which the density maximum of the $\lambda = 66\,\lambda_c$ eigenmode is at $x=1.65$; and 
(ii) Run B (symmetric ICs), which has the density maximum at $x=0$ instead.   
These modes have slightly different amplitudes also, which was necessary to allow both to form clouds due to the faster growth rate of the $\lambda = 66\,\lambda_c$ mode.  Specifically, the amplitude of the $\lambda = 66\,\lambda_c$ mode (with growth rate $n_2$) is set to $A_2 = A_f (A_1/A_f)^{n_2/n_1}$, where $n_1$ and $A_1$ are the growth rate and amplitude of the $\lambda = 132\,\lambda_c$ mode, and $A_f$ is the value of $A_1e^{n_1 t_f} = A_2 e^{n_2 t_f}$ at some time $t_f$ prior to the saturation of TI.  We chose $A_1 = 0.01$ and $A_f = 0.1$, giving $A_2 = 0.0069$.  Under periodic boundary conditions, each setup results in an infinite train of a 4-cloud system once the TI saturates, as shown in the bottom panels of Fig.~1.  

We have explored various other initial conditions including setups using many randomly superimposed entropy modes and setups using random waveforms instead of TI eigenmodes.  
In all cases, the final configuration is either a single cloud or a symmetrically spaced distribution of clouds.  However, Runs A and B are designed to show that a symmetrically spaced configuration of more than one cloud is unstable to small displacements, implying that the only stable configuration is a single cloud system.

The slightly asymmetric ICs of Run~A results in unbalanced pressure forces on the smaller clouds; 
they subsequently acquire a positive (i.e. rightward directed) velocity due to the positive initial displacement.  This is already evident in the left velocity panel of Fig.~1, where we should point out that the larger clouds indeed have larger local velocity fields because they are currently oscillating (see \S{1}).  The velocity fields are directed locally inward in the frame of a given cloud, indicative of a small amount mass advection through the interfaces continuously taking place.  The magnitude of the velocity fields diminish as the oscillations of the clouds damp, as shown in the right velocity panel, yet the smaller clouds in Run~A retain a net positive velocity driven by the pressure gradients (the slopes of the profiles in the bottom right panel).  This net velocity is superimposed on the advective velocity fields that continue to supply (a now tiny amount of) mass. 
  
To see that this is a runaway process, notice that the smaller clouds in Run~B are located at inflection points where the density would reach a minimum and where the velocity field would equal zero if the small clouds were not there.  By displacing the smaller clouds to the right, they basically acquire the nonzero velocity of the local advective component of the velocity field feeding the larger clouds.  Because these velocity fields increase monotonically until reaching the cloud interfaces, the smaller clouds will be swept into the larger clouds at a rate that increases with time. 
We call this the `piggy backing effect' because in the case of clouds with much larger size contrasts than those shown, the velocity profiles of the smaller clouds are effectively perturbations within those of the larger clouds.  We have checked, however, that interacting clouds do not need to be different sizes for coalescence to occur.  The size difference just implies a higher rate of coalescence --- equal size clouds will have oppositely directed advective velocity field components that can more closely cancel each other.  

\begin{figure*}
\includegraphics[width=\textwidth]{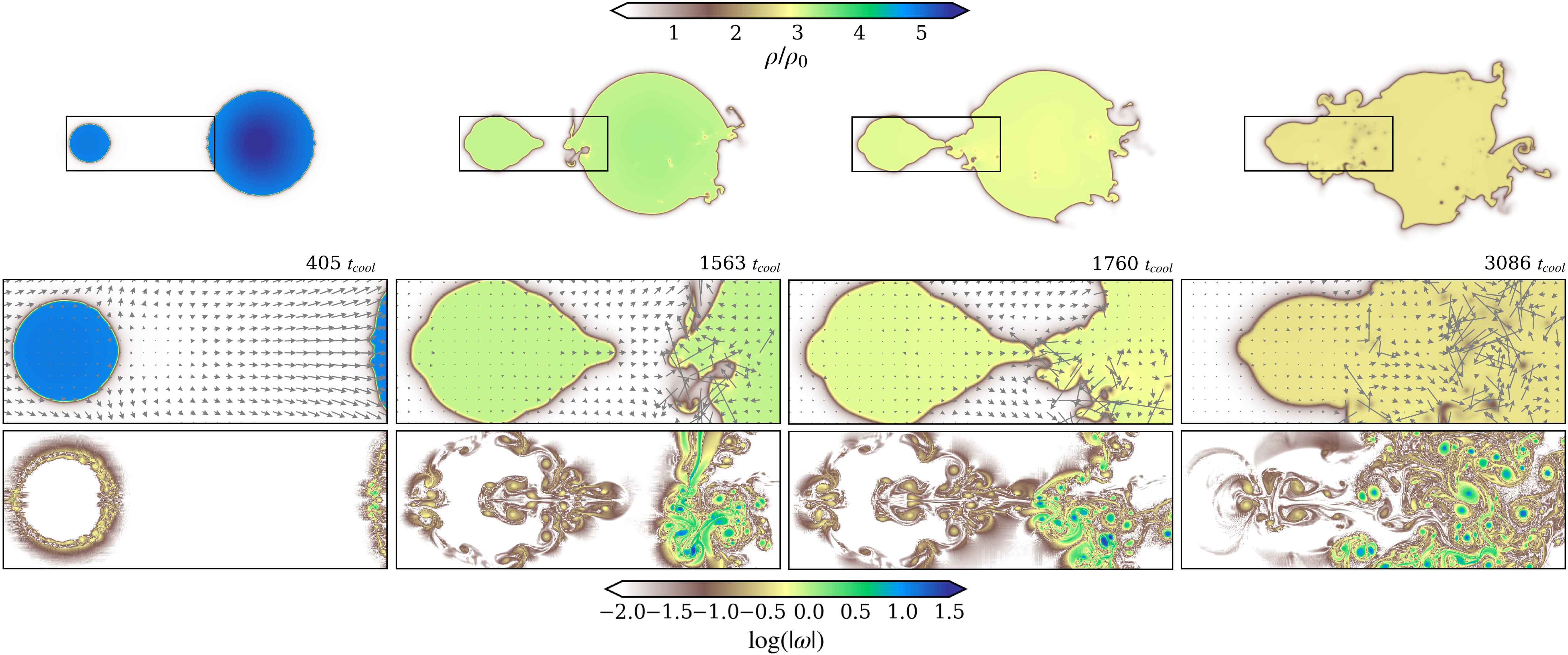}
\caption{Evolution of our fiducial 2D pre-existing cloud simulation.  Top images: 
density colormaps at the four times shown. 
Bottom subpanels: maps of density with velocity vectors overlaid (top) and the vorticity magnitude (bottom; $\omega$ is in units of $c_{s,c}/\lambda_c$) within each rectangular region.  The second panel captures the formation of an entrained vortex bubble.
} 

\end{figure*}

Fig.~2 depicts the further nonlinear evolution of Run~A.   
The lead up to coalescence is an extremely slow process for this particular experiment: it takes $10^4\,t_{cool}$ to reach the state in the first column in Fig.~2.   This is simply a consequence of starting from a state in near-equilibrium; the experiments in \S{3} show that even when starting from stationary clouds, this process can occur at least an order of magnitude faster.
The merger event itself is comparatively rapid, occuring on a dynamical timescale, the final cloud having $t_{dyn} \approx 50\,t_{cool}$.  It induces oscillations (indicated by the `flips' in the pressure profiles), as shown in the next 3 columns.

Notice that the profiles in the first velocity panel no longer possess a visible advective component (characterized by the velocity fields around a given cloud peaking at the interfaces instead of within the core as seen here): these are essentially steady state clouds in relative motion.  
Once they merge, the advective components reappear as the clouds oscillate.
The rightmost panel shows the state after another $\sim 10^4\,t_{cool}$ have elapsed: the tiny advective components of the velocity field are again visible because these two clouds are not in relative motion; correspondingly, the pressure profiles are perfectly symmetric.  We note that this cloud pair can be considered two periods of a 1-cloud system.

\section{Pre-existing cloud simulations}
TI simulations alleviate the need to prescribe adhoc prescriptions for the structure of cloud interfaces, as is common practice (e.g., Nakamura et al. 2006; Pittard \& Parkin 2016; Schneider \& Robertson 2017; Banda-Barrag{\'a}n et al. 2018).  
The non-isobaric regime is computationally expensive to simulate in multi-dimensions, however, so it is desirable to bypass the formation process to enable a more expedient exploration of parameter space.  
To still generate interfaces self-consistently, 
here we `relax' simple round clouds initialized by hand.  The equations we solve and the numerical methods used are the same as in Paper~1, and the resolution is again $16\text{ zones}/\lambda_{c}$.  We apply periodic boundary conditions in both directions.
Cloud interfaces form within just $1\,t_{cool}$ of evolving constant pressure cloud ICs in the presence of conduction and heating and cooling terms.  Specifically, our ICs are $(\rho,\gv{v},p) = (\rho_c/T_c',0,p_0)$ if $r_i < R_i$ and $(\rho,\gv{v},p) = (\chi^{-1}\rho_c/T_c'  ,0,p_0)$ otherwise, where $\rho_c = 2.78 \rho_0$ is the steady state cloud density of our TI runs (see Fig.~2), $\chi=10$ is the initial density contrast,  
$R_i$ is the radius of the $i$-th cloud, and $r_i = \sqrt{(x-x_{c,i})^2 + (y-y_{c,i})^2}$, with $(x_{c,i},y_{c,i})$ the center positions of the $i$-th cloud.  The equilibrium values $(\rho_0, p_0)$ are the same as in Paper~1.
Finally, the parameter $T_c' \equiv T(t=0)/T_c$ sets the initial cloud temperature,
which is chosen to be different from unity (the value for which heating balances cooling) in order to trigger a transient response (see below).

\subsection{2D coalescence dynamics}
\label{2D}
Fig.~3 shows the evolution of our fiducial run: the left and right clouds have initial diameters of $15\,\lambda_c$ and $45\,\lambda_c$, respectively, and we chose $T_c' = 0.8$ (the sensitivity to this parameter is mentioned in \S{3.2}).  
Comparing the first two panels, we see there is an initial decrease in the density and an overall expansion of the clouds, just as in the 1D TI simulations (see Fig.~1). 
The middle two panels show that coalescence in 2D occurs along the centers of the clouds because these regions are closest.  The bottom subpanels zoom-in on the initial contact region to better display the complicated dynamics accompanying vorticity generation.  
Notice how much time elapses between the final two panels.  
This slow evolution is due the transient response to the initial thermal disturbance having mostly died out after $\sim 10^3\,t_{cool}$.  

\subsubsection{Vortex bubble entrainment}
A new and interesting phenomenon is revealed by this 2D simulation: the entrainment of \emph{vortex bubbles}.  
This only occurs for non-isobaric clouds, as only then can there be strong enough oscillations to excite protrusions at the cloud interfaces.\footnote{The protrusions themselves may simply be Kelvin-Helmholtz instability (KHI), but further analysis is needed to establish this.}  
The warmer interface gas becomes fully entrained in a rollup of cloud gas, thereby forming underdense pockets of swirling gas that can subsequently travel deep into the interior of the cloud (see the small brown structures in the final density map).    
We plan to explore these vortex bubbles further and check their observational consequences using 3D simulations. 

\begin{figure}
\includegraphics[width=0.48\textwidth]{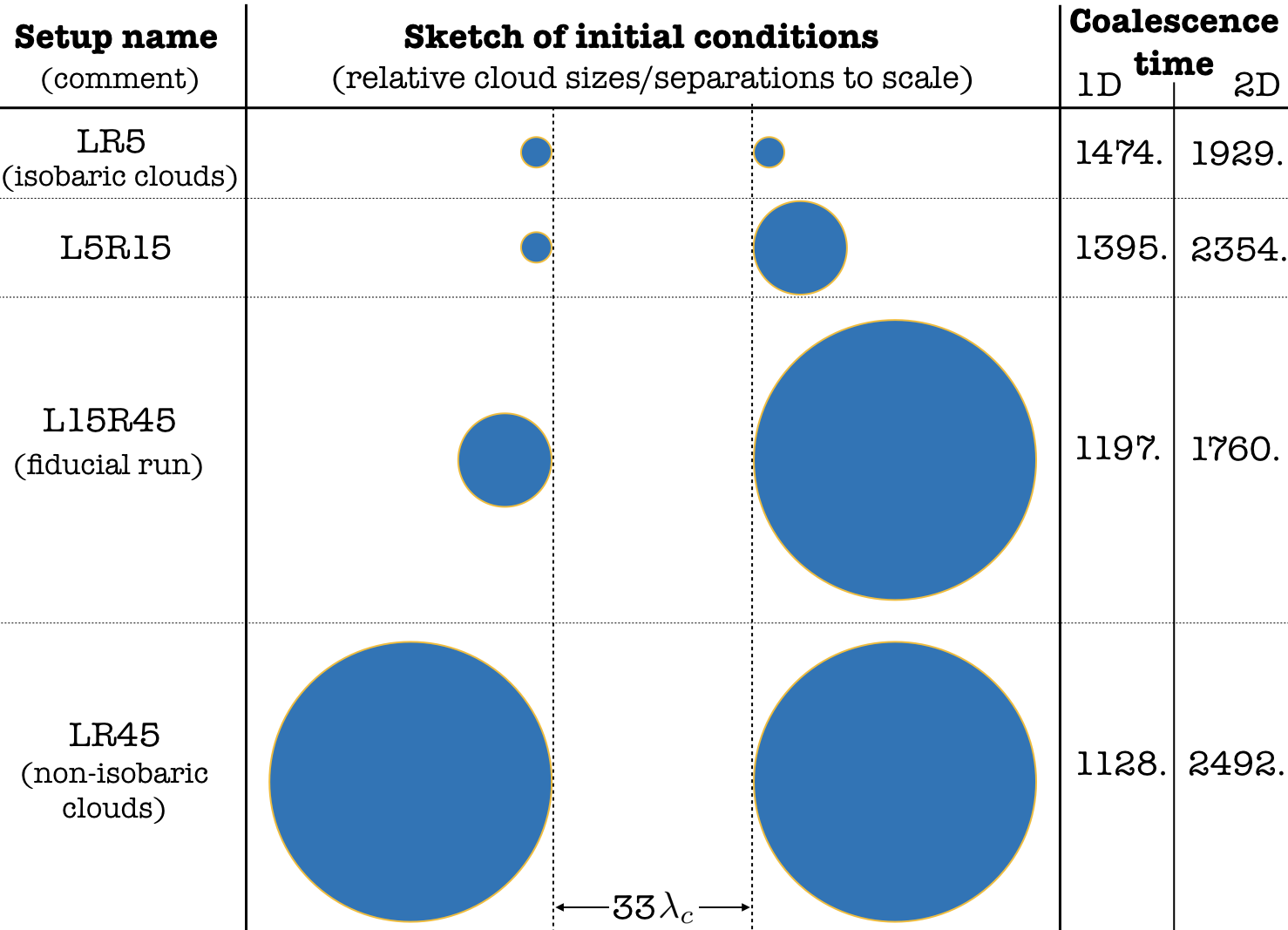}
\caption{Table summarizing our pre-existing cloud setups and the time it takes for each pair of clouds to coalesce (in units of $t_{cool}$) in both 1D and 2D simulations. The times are for an initial contact, not a full merger.  Fig.~3 shows setup L15R45 (with left/right cloud diameters $15\,\lambda_c$/$45\,\lambda_c$).    
}
\end{figure}

\subsection{Factors influencing the coalescence rate}
We have investigated several factors that affect the rate of coalescence including the absolute sizes of the clouds, the contrast in cloud sizes, and the transient response of a cloud to a thermal disturbance. 
The coalescence time will obviously be highly sensitive to the cloud separation distance, but a definitive study of this dependence should assess the competing effects due to intrinsic velocity dispersion. 
For example, for a given cloud distribution, any pair of clouds will have a relative velocity $V$ in general, and for initial trajectories not leading to a collision, we would only expect coalescence to occur if the local advective velocity component surrounding the larger cloud exceeds $V$ at the time of closest approach.  

For a given cloud separation distance, we find that the most important factor determining the coalescence rate is the magnitude of the transient response due to a thermal disturbance.  
Non-isobaric clouds respond to such disturbances by oscillating, giving rise to a transient period during which the magnitude of the local advective velocity is enhanced; its maximum value increases with cloud size (see Paper~1).  
With $T_c' = 1$, coalescence times for pre-existing clouds are comparable to those of the TI simulations from \S{2}, on the order of $10^4\,t_{cool}$ for comparable separation distances.  For $T_c' = 0.8$, meanwhile, the transient response generated reduces the coalescence time an order of magnitude.  

To assess the dependence on the cloud size and size contrasts (for $T_c' = 0.8$ and separation distance $33\,\lambda_c$), we simulated four configurations of two clouds: (1)  LR5 - equal size isobaric; (2) L5R15 - isobaric and non-isobaric; (3) L15R45 - different size non-isobaric; and (4) LR45 - equal size non-isobaric clouds (as schematically summarized in Fig.~4).  
Our fiducial 2D run that we examined above uses setup L15R45.  
The locations of the domain boundaries are always $33\,\lambda_{th}$ from the cloud edges in the $x$-direction and $25\,\lambda_{th}$ beyond the edges of the largest cloud in the $y$-direction.  
We also ran 1D versions of these simulations, in which we evolve profiles given by horizontal cuts through the center of the clouds.  These correspond to `slabs', not round clouds.  Thus, the expectation is that coalescence takes longer in our 2D runs because the separation distance of two round clouds is larger than that of two slabs except along the line through their centers.
 
Fig.~4 compares coalescence times for 1D and 2D runs for each configuration, confirming this expectation.  As already stressed, TI simulations reveal that larger clouds require larger advective velocity field components to maintain their structure as the clouds oscillate.  In consideration of our results from \S{2}, this leads to the expectation that a larger contrast in cloud sizes leads to faster coalescence.  Also, the shortest coalescence times should accompany the largest cloud pairs for clouds of similar size.  Fig.~4 shows that this is indeed the case in 1D, although the differences are minor compared to the effect of varying $T_c'$.  In 2D, meanwhile, we find that coalescence occurs slower for two equally sized non-isobaric clouds compared to two equally sized isobaric clouds.  
 
While it is beyond the scope of this letter, it is likely the case that the coalescence times can be reduced another order of magnitude when the clouds are subject to continual thermal disturbances.  
If the variability timescales of the radiation environment are comparable to $t_{cool}$, non-isobaric clouds will continually oscillate in response; coalescence may then occur on dynamical timescales.

\section{Discussion and conclusions}
The simple numerical experiments presented here have established cloud coalescence as a dynamical instability. 
Our simulations include the necessary physics to self-consistently form interfaces between the clouds and their surroundings, namely heating and cooling processes and thermal conduction --- the same physics underlying TI.  We initialized our pre-existing clouds in a thermally stable plasma, so while cloud coalescence cannot be accurately assessed without this conductive interface physics, it is clear that TI plays no role in this process.

Given our results, what are we to make of a recent numerical study (Sparre et al. 2019; hereafter S+19) that purportedly lends support the shattering hypothesis of M+18 discussed in \S{1}?  A careful examination of this paper reveals that their results are actually consistent with ours.  Since the actual physical mechanism that could trigger a shattering event was not identified by M+18, let us first distinguish between two possible scenarios: (i) shattering is a physical stage in the cloud formation process, namely a nonlinear outcome of TI that affects large perturbations; or (ii) shattering is a phenomenological description of the possible outcome of a large cloud that gets disrupted in some manner.  
In Paper~1, we ruled out the first possibility.  

S+19 explored the second possibility in the context of hot, diffuse galactic outflows by embedding a large cloud in a wind.  
They interpreted the subsequent destruction of the cloud as evidence in support of M+18's shattering hypothesis.  
However, there are a couple inconsistencies in this interpretation.  Most glaringly, the cloudlets in these simulations all first appear at the edges of the large clouds, indicative of shredding due to KHI, not shattering, which M+18 depicted as a process that would uniformly turn the entire cloud into many cloudlets (see fig.~3 of M+18).  
Secondly, S+19, whose simulations did not include thermal conduction, use the friends of friends (FOF) clump finding algorithm to quantify the increase in cloudlet number as the wind shreds the surfaces of their clouds.  If these cloudlets are prone to coalescence, the FOF clump count should decrease at late times, which is indeed evident from their fig.~6, although this was not discussed.  We note that even simulations without thermal conduction should still show some coalescence due to the existence of a numerical Field length (Gazol et al. 2005).  

Finally, other authors have previously recognized the tendency for clouds to coalesce (e.g., Koyama \& Inutsuka 2004).
In particular, even M+18 noted this effect occurring in their most resolved simulation (see their fig.~4), which they referred to as coagulation.  
We would not have expected this effect to be so pronounced in their simulations, considering that their density contrasts are $\chi = 10^3$ (as are S+19's).  
Our simulations have $\chi = 10$, so the inertia of the cold gas is not an important factor in setting the coalescence rate, but for $\chi = 10^3$ it should be (at least until density scales out due to very efficient cooling), and thus clouds should merge on slower timescales than seen here.  In any case,  
M+18 suspected coalescence to be an artifact of their simplified setup and 
further pointed out that turbulence will likely suppress this tendency, which is true (recall \S{3.2}) but beside the point: in controlled experiments that isolate the dynamics of multiple condensations, coalescence occurs.

\acknowledgments
This work was supported by NASA under ATP grant 80NSSC18K1011.
TW is partially supported by the LANL LDRD Exploratory Research Grant 20170317ER.

\end{document}